\documentstyle[prl,aps,epsfig,multicol]{revtex} 
\begin{document} 
\title{The key-lock mechanism in nematic colloidal dispersions} 
 
\author{N.M. Silvestre$^1$, P. Patr\'\i cio$^{1,2}$, and M.M. Telo da Gama$^1$} 
\address{$^1$Departamento de F{\'\i}sica da Faculdade de Ci{\^e}ncias and 
Centro de F{\'\i}sica Te\'orica e Computacional\\ 
Universidade de Lisboa, 
Avenida Professor Gama Pinto 2, P-1649-003 Lisboa Codex, Portugal\\ 
$^2$Instituto Superior de Engenharia de Lisboa\\ 
Rua Conselheiro Em\'\i dio Navarro 1, P-1949-014 Lisboa, Portugal 
} 
 
\date{December 2003} 
 
\maketitle 
 
\begin{abstract} 
We consider the interaction between two-dimensional nematic colloids and 
planar or sculpted walls. 
The elastic interaction between colloidal disks and 
flat walls, with homeotropic boundary conditions, are always repulsive. 
These repulsions may be turned into strong attractions at structured or 
sculpted walls, with cavities that match closely the shape and size of the 
colloids. This 
key-lock mechanism is analyzed in detail for colloidal 
disks and spherocylindrical 
cavities of various length to depth ratios, by minimizing the 
Landau-de Gennes free energy functional of the nematic orientational order 
parameter. 
We find that the attractions occur only for walls with cavities 
within a small range of the colloidal size and a narrow range of 
orientations with respect to the cavity's symmetry axis. 
\end{abstract} 
 
\vspace{2mm} 
PACS numbers:82.70.Dd, 61.30.Hn, 61.30.Jf.
 
\begin{multicols}{2} 
\section{Introduction} 
\label{intro} 
 
In the last ten years, there has been continued interest in colloidal 
dispersions in nematics and other liquid crystalline phases (LCs) 
owing to their novel, complex behavior \cite{Stark_review}. 
The behavior of spherical isotropic particles in a nematic 
matrix depends upon (i) the elastic constants of the nematic, 
(ii) the size of the particle, and (iii) the boundary conditions at the 
particle and at the container, including the anchoring energy of the 
nematogenic molecules and possibly additional (generic) surface tension 
effects. 
All of these contributions are temperature dependent and their combination 
leads to complex anisotropic long-ranged colloidal interactions 
\cite{Ramaswamy_int,Poulin_int,Lubensky_int}. 
These were reported to lead to a variety of novel self-organized colloidal 
structures, such as linear chains \cite{Poulin_sci,Loudet_nat}, 
periodic lattices \cite{Nazarenko_per}, 
anisotropic clusters \cite{Poulin_ani}, and cellular structures 
\cite{Meeker_cel} 
that are stabilized, in general, by topological defects. 
 
More recently, two-dimensional (2D) inverted nematic emulsions were 
also studied and similar behavior has been found 
\cite{Pettey,Cluzeau_1,Cluzeau_2,Patricio,Tasinkevych}. 
In particular, Landau-de Gennes (LdG) theory predicts that the stable 
configuration for a colloidal disk, with strong homeotropic anchoring, is 
always a pair of $1/2$ charge topological defects \cite{Fukuda} and 
thus the long-range interaction between 2D colloids is 
quadrupolar for any sized particles \cite{Tasinkevych}. 
 
The interactions between colloids and the nematic-isotropic (NI) 
interface were also investigated~\cite{West,Andrienko}. These 
include specific contributions from the liquid crystal matrix due 
to distortion of the director field close to the particles and/or 
the interface, and a generic contribution due to wetting and 
surface tension effects. At equilibrium, a strong distortion of the 
planar interfacial region was observed, with the interface bending around 
to wrap the colloid. The 
effective colloid-interface 
interaction was found to be rather complex and a simple scaling 
analysis for a flat interface was shown to fail 
badly~\cite{Andrienko}. 
 
At the same time technological advances allowed the controlled 
fabrication of micro-patterned and structured surfaces, on the nanometer 
to the micrometer scales \cite{Micro_patt}. 
The interplay between surface geometry and orientational order required to 
understand the phenomenology of LCs on substrates patterned on these 
length scales, is however largely unexplored. 
In this article we embark on a systematic investigation of the effects of 
geometry and orientational order on the interaction between colloids and solid 
surfaces. In particular, we investigate the interaction between 2D colloids 
and hard surfaces, ranging from flat to structured on the colloidal scale. 
We study the interaction between one disk and flat as well as structured 
walls, using the method of images and numerical minimization the LdG 
free energy. For a single cavity, sculpted on a flat 
surface, we find that the flat wall repulsion may be turned into a 
rather strong 
attraction, as a result of the interplay between geometry and orientational 
order. This effect occurs at cavities with sizes in the colloidal range 
resembling the so-called {\it key-lock} mechanism. 

In section II, we review briefly the Landau-de Gennes 
theory and in section III we present the results for the interaction 
between one disk and a flat wall. In section 
\ref{patterned} we consider the interaction of colloidal 
disks with cavities sculpted on flat walls. We investigate the effects of 
the size and shape of the cavity and of the colloidal orientation with 
respect to the cavity's symmetry axis. 
Finally, in section V we summarize our results and make some concluding 
remarks. 
 
\section{Landau-de Gennes functional} 
\label{model} 
 
In what follows we consider a two-dimensional nematic liquid crystal. 
On average, the molecules are aligned along one common direction 
described by the director {\bf n} and the tensor order parameter 
is defined as 
$Q_{\alpha\beta}({\bf r})=Q({\bf r})(n_\alpha ({\bf r})n_\beta ({\bf r}) 
-\delta_{\alpha\beta}/2)$ \cite{deGennes}. In the situations to be 
investigated, the distances over which significant variations of 
$Q_{\alpha\beta}$ occur are much larger than molecular dimensions. 
Thus density variations are neglected. The free energy density is then 
written in terms of invariants of {\bf Q} and its derivatives and is 
known as the Landau-de Gennes free energy, 
\begin{eqnarray} 
F[Q_{\alpha\beta},\nabla Q_{\alpha\beta}]=\int_\Omega d^2r\Big[&-&\frac{A}{2} 
{\it Tr}\{{\bf Q}^2\}+\frac{C}{4} {\it Tr}\{{\bf Q}^2\}^2 \nonumber \\ 
&+&\frac{L}{2} \nabla_\gamma Q_{\alpha\beta}\nabla^\gamma Q^{\alpha\beta} \Big], 
\label{Landau_deGennes} 
\end{eqnarray} 
where we used the one elastic constant approximation. {\it Tr} 
denotes the trace operation, $\Omega$ is the area of the 2D system, 
$A$ and $C$ are bulk constants and $L$ is the elastic constant. 
Stability requires $C>0$. In the nematic phase $A>0$ and the equilibrium 
orientational order parameter is $Q_{bulk}=\sqrt{2A/C}$. For simplicity, we 
consider
strong homeotropic (perpendicular) anchoring meaning that the order 
parameter 
at the wall and disk boundary is fixed and equal to $Q_{bulk}$. 
The colloid is modeled as a hard disk, 
and all other colloidal interactions (van der Waals, electrostatic, etc.) 
are neglected.
Thermal fluctuations, also neglected within the Landau-de Gennes 
description, are 
expected to 
be unimportant for the systems considered here: hard surfaces at 
temperatures well below the NI transition temperature. 
 
We use finite elements with adaptive meshing, as described in 
\cite{Patricio}, to minimize the Landau-de Gennes free energy functional. 
Indeed, the major difficulty in the numerical problem stems from the 
widely different length scales set by the disk or the sculpted wall and 
the defects. 
A first triangulation respecting the (predefined) geometrical boundaries 
is constructed. 
The tensor order parameter $Q_{\alpha\beta}({\bf r})$ is given at the 
vertices of this mesh and linearly interpolated within each triangle. The 
free energy is then minimized using standard methods. 
The variation of the solution at each iteration is 
used to generate a new adapted mesh. In the far-field  
$Q_{\alpha\beta}({\bf r})$ varies slowly and the triangles are large. By 
contrast, 
close to the defects the tensor order parameter varies rapidly  
and the triangles are several orders of magnitude smaller. In 
typical calculations convergence is obtained after two mesh adaptations, 
corresponding to final 
meshes with $10^4$ points, spanning a region of $20{\it a}\times 
20{\it a}$, and minimal mesh sizes of $10^{-4}{\it a}$, close to the 
defects. The corresponding free energy is given with a relative 
accuracy of $10^{-4}$. 
 
\section{Flat wall} 
\label{planar} 
 
We start by considering the interaction of a colloidal disk, of radius 
$a$, with a flat wall. We assume strong homeotropic boundary 
conditions at the wall and at the disk's surface, and take the 
far field ${\bf n}_0$ perpendicular to the wall. The order parameter at 
the wall and at the disk's surface is fixed and is equal to $Q_{bulk}$. 
 
The fixed homeotropic anchoring at the boundaries reduces the 
(long-range) flat disk-wall interaction to the interaction between two 
disks, studied in Ref.\cite{Tasinkevych}. The latter was
obtained using an electromagnetic analogy \cite{Pettey} establishing that, 
at large separations, each disk is accompanied by a pair of defects and 
behaves as a quadrupole. 
This still applies to the flat wall system with the result that 
the 
long-range disk-wall interaction is repulsive and decays as $R^{-4}$, where 
$R$ is the 
distance between the center of the disk and the wall. 
However, at short range,
nonlinear effects come into play, and the electromagnetic analogy is
no longer valid.
Moreover, in that region the equilibrium disk-disk 
interaction spontaneously breaks the flat wall (mirror) symmetry 
\cite{Tasinkevych} 
and thus it is no longer useful in this problem. 
In this regime, we resorted to numerical solutions. 
 
\begin{figure}[ht] 
\par\columnwidth=20.5pc 
\hsize\columnwidth\global\linewidth\columnwidth 
\displaywidth\columnwidth 
\centerline{\epsfxsize=220pt\epsfbox{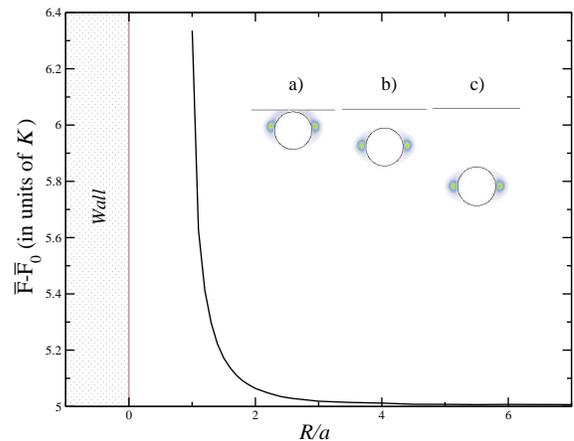}} 
\caption 
{(Color online) 
Reduced interaction energy as a function of the distance $R$ from the 
center of the colloid, of radius $a$, to the wall. ${\bar F}=F/K$, where 
$K=2LQ_{bulk}$ is the Frank elastic constant, and $F_0$ is the 
Landau-de Gennes free energy of the system without colloid. The inset 
illustrates the order parameter maps at different $R$. a) 
$R/a=1.1$; b)$R/a=2.0$; c)$R/a=4.0$. The nematic order 
parameter varies between $Q=Q_{bulk}$ (white regions) and $Q=0$ (colored 
regions).} 
\label{planar_fig} 
\end{figure} 
 
In Fig.\ref{planar_fig} we plot the interaction energy as a function of 
$R$. As noted above, the long-range interaction is always 
repulsive and we find that the repulsion increases as the colloid 
approaches the wall. 
This increased short-range repulsion is associated with the strong distortion 
of the nematic matrix between the (flat) wall and the (curved) disk, due 
to the competition between the fixed anchoring conditions at the wall and 
disk surfaces. 
The distortion of the nematic matrix extends from the wall up to the line 
joining the defects and under these circumstances the elastic free energy is 
minimized if the defects move closer to the wall.
As a result the pair of 1/2 defects surrounding the disk are 
displaced with respect to their (symmetrical) equatorial position in 
the isolated disk. 
This effect is illustrated in the inset of Fig.~1,
exhibiting order parameter maps at different disk-wall separations, 
$R$.
In Fig.~2, we plot the defects orientation, $\delta\theta$,
with respect to their orientation in isolated disks, as a function of $R$ 
(see inset for the notation).
In addition,
the distance of the defects from the center of the disk, 
$r_d/a$,
decreases at small disk-wall separations, $R$,
as the defects change their orientation.
 
\begin{figure}[ht] 
\par\columnwidth=20.5pc 
\hsize\columnwidth\global\linewidth\columnwidth 
\displaywidth\columnwidth 
\centerline{\epsfxsize=220pt\epsfbox{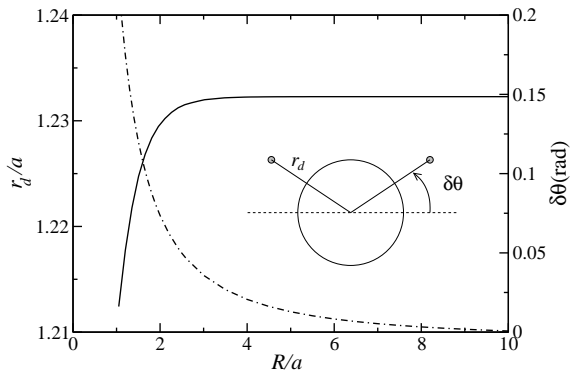}} 
\caption 
{Defect distance $r_d$ from the 
center of the disk (continuous line) and defects orientation 
$\delta\theta$ (dash-dotted line), with respect to their 
orientation in isolated disks, as a function of $R$.} 
\label{deviation} 
\end{figure} 
 
\section{Structured wall} 
\label{patterned} 
 
The results of the previous section indicate that a homeotropic disk is 
repelled by a hard (homeotropic) wall since the elastic deformation  
of the nematic matrix increases as the colloid approaches the wall.
This situation may be inverted if the director field at the wall resembles 
the director field close to an isolated disk. In the 
following we explore this mechanism and determine the conditions under 
which it is strong enough to overcome the flat wall repulsion.

\begin{figure}[ht] 
\par\columnwidth=20.5pc 
\hsize\columnwidth\global\linewidth\columnwidth 
\displaywidth\columnwidth 
\centerline{\epsfxsize=220pt\epsfbox{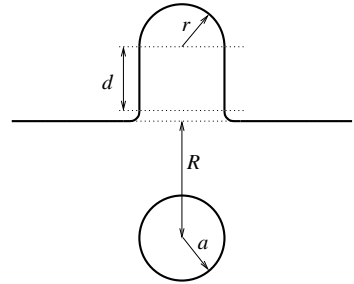}
} 
\caption 
{Structured wall: geometry and notation.} 
\label{patterned_fig} 
\end{figure}

We consider a sculpted cavity on an otherwise flat wall, as shown in 
Fig.~\ref{patterned_fig}. The cavity is spherocylindrical of radius $r$ 
and half-length or depth, $d$. The corners are replaced by an arc of 
circle with radius $a/4$, to avoid large variations in the director 
field, in the absence of the colloid. Strong homeotropic 
boundary conditions are set at the wall and at the colloidal surface. 
The far field ${\bf n}_0$ is again perpendicular to the wall. 

The image method can no longer be applied and analytical 
expressions for the long-range interaction between the disk and the 
cavity are not available in this case. In addition, the non-linearities 
that come into 
play, as the colloid approaches the cavity, are more complex. 
The free energy is minimized numerically as before, using 
finite elements with adaptive meshing \cite{Patricio}. 

\begin{figure}[ht] 
\par\columnwidth=20.5pc 
\hsize\columnwidth\global\linewidth\columnwidth 
\displaywidth\columnwidth 
\centerline{\epsfxsize=220pt\epsfbox{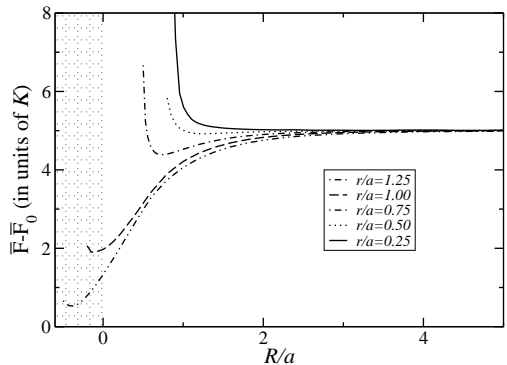}} 
\caption 
{Reduced interaction energy as a function of the distance, $R$, of the 
center of the colloid to the midpoint of the cavity's entrance, for 
several widths ($r/a=0.25$,$ 0.50$,$ 0.75$,$ 1.00$,$ 1.25$). 
The depth is $d/a=0.01$.} 
\label{small_d} 
\end{figure}

We start by considering a very 
shallow cavity, with $d/a=0.01$. 
In Fig.~\ref{small_d} we show the disk-wall interaction 
as a function of the separation $R/a$, for increasing values of 
the cavity's radius. 
If the cavity is sufficiently narrow ($r/a<0.25$) the colloid-wall 
interaction is 
very similar to the interaction with the flat wall. As the radius of the 
cavity increases, 
an attraction with a well defined minimum eventually 
occurs. The minimum is inside the cavity if the width is larger than 
the colloidal radius, $r/a>1$. 

However, if we increase the radius of the cavity well beyond the 
colloidal size, the attraction becomes very weak. When the 
radius of the cavity is much larger than $a$, the 
interaction between the disk and the structured wall approaches that of 
the flat wall. 

\begin{figure}[ht] 
\par\columnwidth=20.5pc 
\hsize\columnwidth\global\linewidth\columnwidth 
\displaywidth\columnwidth 
\centerline{\epsfxsize=240pt\epsfbox{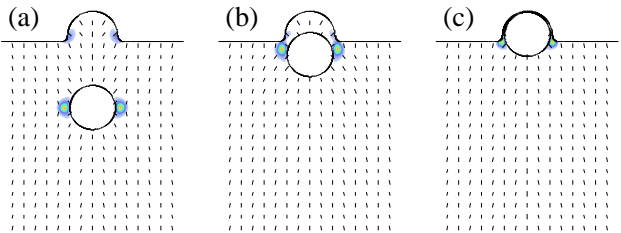}} 
\caption 
{(Color online) 
Order parameter maps for a cavity with $r/a=1.1$ and $d/a=0.01$, 
and different colloidal separations $R$. (a) $R/a=3.0$; 
(b) $R/a=0.6$; (c) $R/a=-0.32$. The nematic order parameter varies 
between $Q=Q_{bulk}$ (white regions) and $Q=0$ (colored regions).} 
\label{3_shots} 
\end{figure}

The disk-wall attraction that occurs when the radius of the cavity is of 
the order of the particle radius, is easily understood in terms of the 
nematic 
configurations of Fig.~\ref{3_shots}, where we plot the nematic order 
parameter maps and the director orientation, for a cavity 
with $r/a=1.1$ and $d/a=0.01$, and different colloidal 
separations, $R$. The shape of the cavity is 
complementary to that of the disk. The distortion due to the  
homeotropic anchoring at all surfaces is clearly minimized when the 
particle fills the cavity. In fact, as the colloid approaches the 
cavity, the two $1/2$ defects move towards the corners and merge 
with the distorted region of the nematic matrix in the absence of 
the colloid (see Fig.~\ref{3_shots}a). Due to the strong  
distortion of the nematic matrix at the corners of the isolated cavity, 
the tensor order 
parameter decreases in this region, even if we keep 
$Q=Q_{bulk}$ fixed at the surface. 
Different surface interactions will lead to different 
director configurations but for simplicity we restrict the study to 
strong anchoring conditions.

\begin{figure}[ht] 
\par\columnwidth=20.5pc 
\hsize\columnwidth\global\linewidth\columnwidth 
\displaywidth\columnwidth 
\centerline{\epsfxsize=220pt\epsfbox{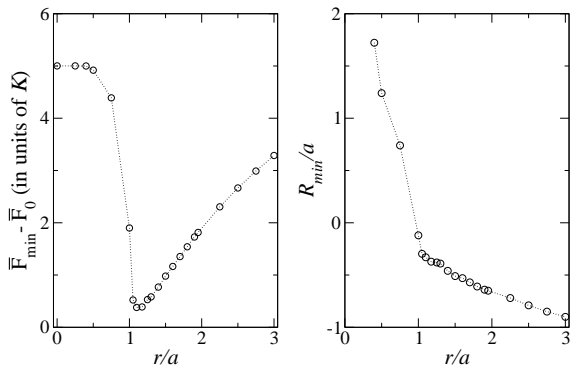}} 
\caption 
{Minimal interaction energy (left) and equilibrium position $R_{min}$ 
(right) 
as functions of the radius of the cavity $r/a$.} 
\label{fig6} 
\end{figure} 

In Fig.~\ref{fig6} we plot the minimal interaction 
energy, $\bar{F}_{min}-\bar{F}_0$ (left), and the corresponding 
disk-wall separation, $R_{min}/a$ (right), as a function of the radius of 
the cavity, $r/a$. 
For small $r/a$, the disk-wall interaction is repulsive,
and the minimal free energy occurs when the colloidal particle
is far from the wall, $R\to\infty$.
The disk-wall interaction becomes attractive at 
finite $R_{min}/a$ (corresponding to a critical 
cavity radius $r_c/a \simeq 0.4$). 
Beyond this point, the minimal free energy decreases rapidly
and reaches its lowest value for a cavity with $r_m/a\simeq 1.1$.
The strongest disk-wall attraction occurs for this geometry, and is 
illustrated in Fig.\ref{3_shots}.
 
\begin{figure}[ht] 
\par\columnwidth=20.5pc 
\hsize\columnwidth\global\linewidth\columnwidth 
\displaywidth\columnwidth 
\centerline{\epsfxsize=220pt\epsfbox{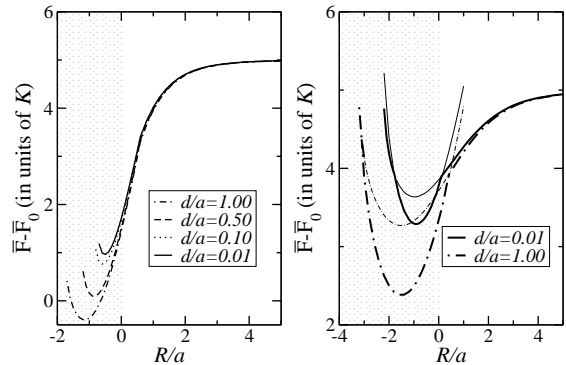}} 
\caption { Left: interaction energy as a function of the separation $R$, 
for 
several cavities ($d/a=0.01$, $0.10$, $0.50$, $1.00$) with radius 
$r/a=1.5$. 
Right: interaction energy as a 
function of the separation $R$, for several depths ($d/a=0.01$,$1.00$). 
The radius of the cavity is $r/a=3.0$. The lines in bold correspond to 
the minimal energy configurations. Thinner lines correspond to metastable solutions.} 
\label{fig7} 
\end{figure}

In the following we show that the disk-wall 
attraction, discussed for shallow cavities, may be enhanced by 
increasing the cavity depth, $d$. In 
Fig.\ref{fig7} we plot 
the reduced interaction energy as a function of the separation $R$, 
for cavities with different depths, $d$. The cavity radius is  
$r/a=1.5$ and $r/a=3.0$, on the left and right of the figure, 
respectively. Increasing $d$ reduces the equilibrium  
interaction energy by pulling the disk deeper into the cavity. 
The nematic distortion is decreased and the defects are further localized 
at the corners leading to a significant reduction in the elastic free energy. 
For cavities with $r/a=3.0$ the analysis revealed the existence of 
a second (metastable) solution where the defects remain attached to the 
colloid.
This is illustrated in Fig.\ref{fig7} (right) 
where we plot the energy of both configurations for two different 
cavities, $d/a=0.01$ 
(full lines) and $d/a=1.00$ (dashed lines). 
The lines in bold correspond to the minimal energy branches.

When $r\sim a$, the disk-wall attraction is enhanced as the depth, 
$d$, increases. A similar effect occurs for wider cavities.
However, as was pointed out before, the attraction becomes weaker, 
and eventually, turns into a repulsion as the wall becomes flatter. 
 
\begin{figure}[ht] 
\par\columnwidth=20.5pc 
\hsize\columnwidth\global\linewidth\columnwidth 
\displaywidth\columnwidth 
\centerline{\epsfxsize=220pt\epsfbox{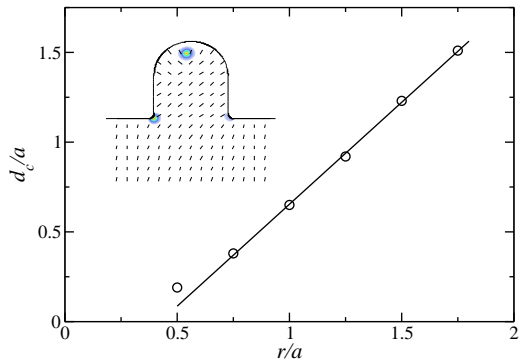}} 
\caption {(Color online)
Critical depth $d_c$ as a function of the radius of the cavity.
The continuous line is the fit $d_c/a=0.48+1.14 r/a$.
Inset: Order parameter map for a nematic filled cavity of depth 
$d\gg d_c$. The nematic order parameter varies between 
$Q=Q_{bulk}$ (white regions) and $Q=0$ (colored regions).} 
\label{fig8} 
\end{figure}

The enhancement of the disk-wall attraction occurs only for a 
limited range of depths. At a certain depth the nematic 
configuration changes abruptly. In fact, due to the strong 
homeotropic boundary conditions, there is a critical depth, $d_c$, 
beyond which the stable nematic configuration of the empty cavity 
exhibits two topological defects (see inset of Fig.~\ref{fig8}). 
One topological defect is inside the cavity, near the cap, while the other is 
pinned near one of the corners. This broken symmetry configuration 
is two-fold degenerate. Along the cavity's neck the nematic 
director is almost constant and this configuration will be 
badly distorted by a colloidal disk. This results in a free 
energy barrier leading to a strong colloidal repulsion. The barrier 
will vanish 
for cavities that are wide enough to allow a smooth 
deformation of the director field when the colloidal particle is 
inserted. 
  
Simple dimensional arguments may be used to understand this 
configurational instability.
For smooth deformations, the elastic free energy of the 
nematic is proportional to $Kd/r$. 
By contrast, the free energy of a uniform nematic aligned along the 
cavity's neck, with two 1/2 defects placed symmetrically at both ends,
scales as, $\pi q^2K$, the core energy of the defects.
Thus, the latter is the equilibrium configuration for sufficiently large 
$d$.
However, as seen in the inset, at the instability a symmetry breaking 
transition also occurs and the lower defect is pinned near one corner.
In order to ascertain the validity of the simple dimensional arguments we 
have checked numerically that the energy of the smooth deformation scales 
with $d$, while that of the broken symmetry configuration is constant.
The critical depths $d_c/a$ were 
also calculated numerically and are plotted in Fig.~\ref{fig8} as a 
function 
of the cavity radius, $r/a$.
At large $r/a$ this exhibits the linear relation, $d_c/a=0.48+1.14 
r/a$.
 
\begin{figure}[ht] 
\par\columnwidth=20.5pc 
\hsize\columnwidth\global\linewidth\columnwidth 
\displaywidth\columnwidth 
\centerline{\epsfxsize=220pt\epsfbox{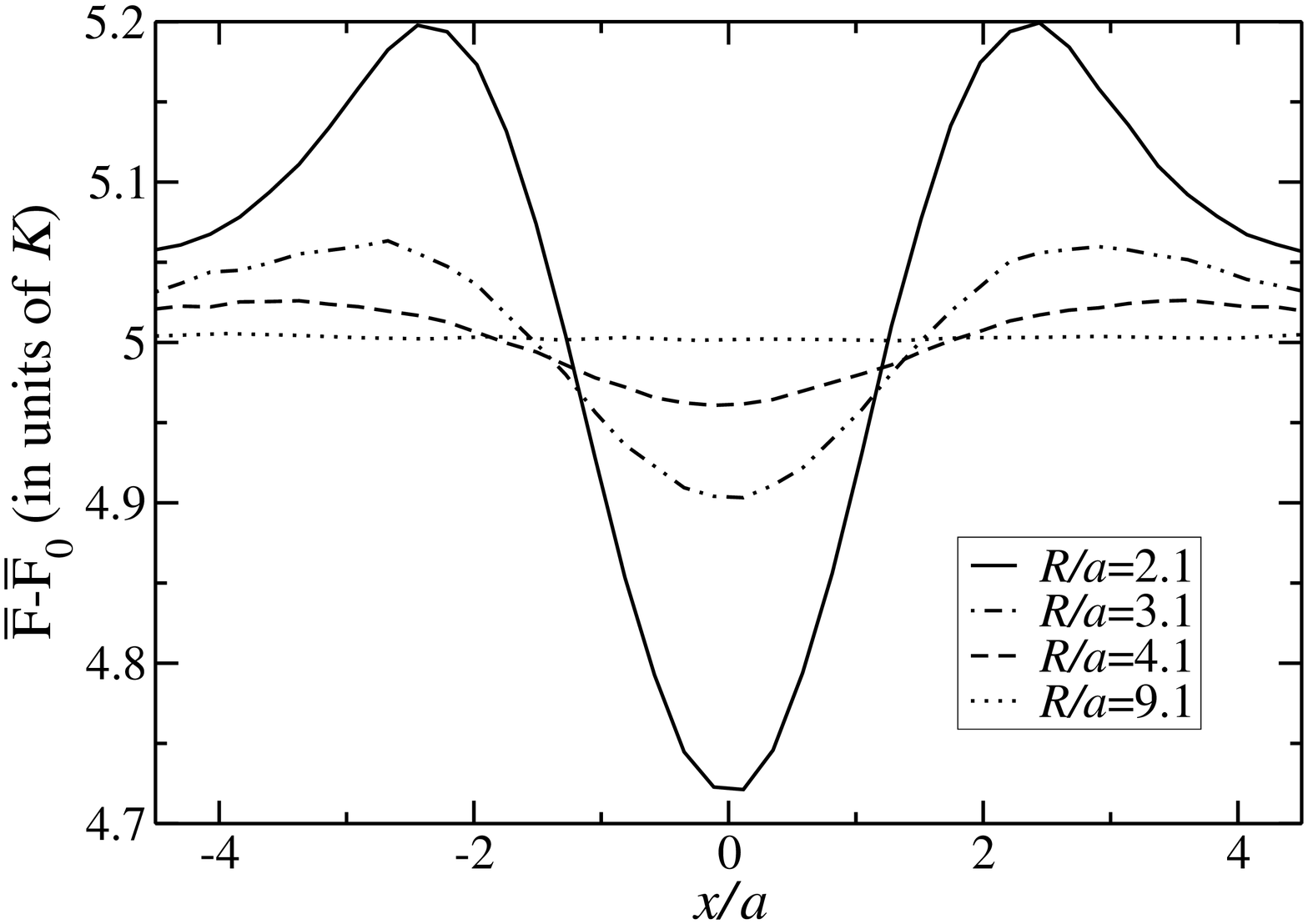}} \caption {Reduced 
interaction energy along four lines parallel to the wall 
($R/a=2.1$, $3.1$, $4.1$, $9.1$) as a function of the lateral 
distance from the cavity, $x/a$. The radius and depth of the 
cavity are $r/a=1.5$ and $d/a=1.00$, respectively. } 
\label{4_paths} 
\end{figure}

The results so far considered the interaction between colloidal 
disks and sculpted walls along the symmetry axis of the cavity 
(see Fig.~\ref{patterned_fig}). In the remaining of this section we 
consider the interaction energy along different directions. In 
Fig.~\ref{4_paths} we plot the interaction energy along lines 
parallel to the wall ($R/a=2.1$, $3.1$, $4.1$, $9.1$), as a 
function of the lateral distance from the cavity, $x/a$, for a 
cavity with $r/a=1.5$ and $d/a=1.00$. Note that the attraction is 
limited to a certain angle 
that depends on the colloidal separation, $R$. Outside that 
region, the colloid is repelled by the flat wall and by the 
cavity corners. Indeed, a strong variation of the interaction 
energy as the disk approaches the corners may be seen in 
Fig.~\ref{4_paths}. 
 
The colloid will be trapped by an appropriately sized cavity if, 
and only if, it is (driven) inside the cavity's 'cone' of attraction. 
 
\section{Conclusions} 
\label{conclude} 
 
We studied the interaction of a colloidal disk, in a 2D nematic, with a 
flat wall with strong homeotropic boundary conditions and found that it is 
purely 
repulsive and decays at long-range as $R^{-4}$. 
By sculpting the surface with a cavity that is similar in size and shape 
to the colloid we found a robust key-lock mechanism, capable of turning 
the repulsion into an attraction large enough to trap 
colloidal particles. 
 
This key-lock mechanism is clearly dependent on the geometry of 
both the colloid and the cavity and its effectiveness for other 
colloidal shapes will be addressed in future work. 
Another relevant question that is left open is the existence of a similar 
effect at soft (deformable) walls. This will allow contact with recent 
results at NI interfaces where colloids were found to be trapped through 
a mechanism that is somewhat more complex, since the bending of the 
interface results from the colloidal interactions \cite{Andrienko}. 
 
Lastly, we note that the key-lock mechanism has features reminiscent of 
the process of wrapping colloids by a 
membrane~\cite{Boulbitch,Dezerno}. It is likely that 
the major factors determining this wrapping process are most simply 
illustrated for the type of hard structured walls considered in this work. 
However, a detailed comparison of hard and soft surfaces requires further 
analysis based on the LdG approach and/or effective Hamiltonian models, in 
order to assess (among other things) the role of thermal fluctuations. 
 
We end with the remark that our results for the 2D key-lock 
mechanism are applicable to a particular 3D system, consisting of long 
rod-like colloids with their major axes parallel to a planar surface. 
Methods to manipulate such systems have already been developed \cite{Leheny}.
Three 
dimensional effects, such as the biaxiality at the defect cores, are 
small~\cite{Andrienko_Tasinkevych} and thus the key-lock 
mechanism reported here should be valid for these 3D systems where the 
non-uniformity is (quasi) two-dimensional.

\end{multicols} 
 
\end{document}